\documentclass[a4paper,11pt]{article}
\usepackage{pos}
\usepackage{wrapfig}   

\title{First IACT Waveform Analysis Based on Deep Convolutional Neural Networks Using CTLearn}
\ShortTitle{First CNN-based IACT Waveform Analysis using CTLearn}

\manuallySeparateAuthors
\author*[a]{T. Miener}
\author[a]{, L. Burmistrov}
\author[a]{, B. Lacave}
\author[b]{, and A. Cerviño}
\author[c]{ for the CTAO-LST project}
\affiliation[a]{Département de Physique Nucléaire et Corpusculaire, Université de Genève, Faculté de Sciences,\\
1211 Genève 4, Switzerland}
\affiliation[b]{IPARCOS-UCM, Instituto de Física de Partículas y del Cosmos, and EMFTEL Department, Universidad Complutense de Madrid, Plaza de Ciencias, 1. Ciudad Universitaria, 28040 Madrid, Spain}
\affiliation[c]{\href{https://www.lst1.iac.es}{https://www.lst1.iac.es}; see full author list of the CTAO-LST Project at the end of the document}

\emailAdd{tjark.miener@unige.ch}

\abstract{Imaging atmospheric Cherenkov telescopes (IACTs) detect extended air showers (EASs) generated when very-high-energy (VHE) gamma rays or cosmic rays interact with the Earth's atmosphere. Cherenkov photons produced during an EAS are captured by fast-imaging cameras, which record both the spatial and temporal development of the shower, along with calorimetric data. By analyzing these recordings, the properties of the original VHE particle—such as its type, energy, and direction of arrival—can be reconstructed through machine learning techniques. This contribution focuses on the Large-Sized Telescopes (LSTs) of the Cherenkov Telescope Array Observatory, a next-generation ground-based gamma-ray observatory. LSTs are responsible for reconstructing lower-energy gamma rays in the tens of GeV range. We explore a novel event reconstruction technique based on deep convolutional neural networks (CNNs) applied on calibrated and cleaned waveforms of the IACT camera pixels using CTLearn. Our approach explicitly incorporates the time development of the shower, enabling a more accurate reconstruction of the event. This method eliminates the need for charge integration or handcrafted feature extraction, allowing the model to directly learn from waveform data.}

\ConferenceLogo{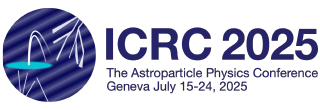}

\FullConference{39th International Cosmic Ray Conference (ICRC2025)\\
15–24 July 2025\\
Geneva, Switzerland\\}

\begin{document}
\maketitle

\section{Introduction}
Very-high-energy (VHE; approximately above 20 GeV) gamma-ray astronomy enables the study of some of the most extreme and energetic phenomena in the universe, such as pulsars, active galactic nuclei, and gamma-ray bursts. Ground-based detection of these gamma rays is made possible by Imaging Atmospheric Cherenkov Telescopes (IACTs), which observe the Cherenkov radiation emitted by extensive air showers (EASs) initiated when VHE particles interact with the Earth's atmosphere. The Cherenkov photons, typically in the near-ultraviolet to visible range, are collected by segmented mirrors and imaged onto fast cameras capable of capturing both the spatial distribution and temporal evolution of the shower.

The reconstruction of the primary particle's properties, such as its type, energy, and arrival direction, from these recordings is a central challenge in IACT data analysis. Traditional approaches rely on charge integration over the waveform, followed by handcrafted feature extraction, such as Hillas parameters~\cite{1985ICRC....3..445H}, and classification or regression models built on these features~\cite{2008NIMPA.588..424A}. While effective, these methods can be limited by their reliance on prior assumptions and may discard potentially informative temporal data embedded in the calibrated waveforms.

The Cherenkov Telescope Array Observatory (CTAO)~\cite{2019APh...111...35A}, currently under construction, represents the next-generation ground-based observatory for very-high-energy (VHE) gamma rays. With sites in both the northern and southern hemispheres, CTAO will provide full-sky coverage and improved operational duty cycle. It will employ a hybrid array of telescopes of different sizes to maximize sensitivity across the VHE gamma-ray spectrum. The Large-Sized Telescopes (LSTs), with their large mirror area and fast electronics, are optimized for detecting low-energy gamma rays in the tens of GeV range, where precise waveform processing is especially critical for accurate event reconstruction.

Advances in machine learning (ML) over the last decades, particularly deep learning (DL) with convolutional neural networks (CNNs), offer an opportunity to revisit the reconstruction problem by allowing models to learn directly from minimally processed data. In this work, we present a novel event reconstruction approach for the LSTs using CTLearn\footnote{\href{https://github.com/ctlearn-project/ctlearn}{https://github.com/ctlearn-project/ctlearn}}~\cite{miener_2025_15065761,Nieto:2019ak}, a DL framework developed for IACT data analysis. Our method applies CNNs to calibrated and cleaned waveforms from the IACT camera pixels, preserving the full temporal structure of the recorded signals. By avoiding intermediate charge integration and feature engineering steps, the model learns to extract discriminative features directly from the waveform data, enabling improved reconstruction of particle energy and direction.

\section{Data analysis pipeline}

The analysis is carried out using a simulated, mono-telescope dataset specifically tailored to the LST-1 prototype and its observational configuration. This dataset was generated with \texttt{CORSIKA}~\cite{Heck:1998vt} and \texttt{sim\_telarray}~\cite{Bernlohr:2008kv}, following the standard CTAO simulation procedures. The raw simulation data were processed into calibrated waveforms and cleaned images using \texttt{ctapipe}~\cite{karl_kosack_2025_15606984}, the prototype low-level data processing pipeline for CTAO. The resulting dataset conforms to the \texttt{ctapipe} reference implementation of the CTAO data format and includes calibrated waveforms for each event. This specific simulated dataset was also employed in the LST-1 performance study~\cite{Abe_2023}.

To make the training of the DL models feasible, we restricted the dataset to a narrow range of observation altitudes. Specifically, we used for the training of the models only two adjacent pointing nodes, following the strategy of~\cite{Lacave:2025}. For the gamma-diffuse simulations, these correspond to zenith angles of $ 9.579^{\circ} $ and $ 16.087^{\circ} $, and azimuth angles of $ 126.888^{\circ} $ and $ 108.090^{\circ} $, respectively. Due to the limited number of proton events available in each pointing node, we also included the two immediate neighboring pointing nodes, as well as four pointing nodes with the same zenith angles but opposite azimuth directions. This allowed us to construct a more balanced training set between gamma-ray and proton samples. A generic model trained on a wide range of pointing nodes, as done in~\cite{Lacave:2025} using integrated and cleaned images, is not yet feasible with waveform data due to current limitations in hardware capabilities and data throughput.

The DL models were trained using diffuse gamma-ray and proton events, generated within cones of $ 2.5^{\circ} $ and $ 8^{\circ} $ radius, respectively. Regression models for energy and arrival direction reconstruction were trained only on the diffuse gamma-ray training set. Performance evaluation was carried out using independent samples of simulated protons ($ \sim 10^{7} $ events; zenith $ 10.0^{\circ} $; azimuth $ 102.199^{\circ}  $; energy range $ 10~\text{GeV}-103.9~\text{TeV} $) and pointlike gamma rays ($ \sim 10^{7} $ events; zenith $ 10.0^{\circ} $; azimuth $ 102.199^{\circ}  $; energy range $ 5~\text{GeV}-54.5~\text{TeV} $), assuming a gamma-ray point source uniformly from a $ 0.4^{\circ} $ offset of the telescope pointing (ringwobble). Due to the lack of testing protons data, we can only evaluate the classifier model with the signal distribution.

\texttt{CTLearn} was employed for DL-based particle classification and event reconstruction, such as the regression for the energy and the arrival direction, using calibrated waveforms as input. The predictions of \texttt{CTLearn} are stored in \texttt{ctapipe}-compatible format, aligned with the CTAO reference data structure to ensure consistency and interoperability within the Data Processing and Preservation System (DPPS) of CTAO. To ensure reproducibility with minimal manual intervention, we utilized the \texttt{CTLearn-Manager} package, which provides an automated and configurable interface for streamlining DL training, evaluation, and data handling workflows.

\section{Methodology}

We adopt a similar methodological framework as presented in~\cite{Lacave:2025}, with adaptations for the use of calibrated waveforms. Instead of integrated charge and peak arrival time images, we input the calibrated waveforms, cropped for 20 samples of 1 nanosecond per pixel, into the CNN. Although waveform data is used, we still apply the standard image cleaning mask from conventional IACT analyses to remove background-dominated pixels and retain those containing significant Cherenkov signal. This cleaning step improves robustness and maintains applicability to real observational data. In future work, this cleaning step is planned to be replaced by a DBSCAN-based algorithm applied directly to the waveform data (see section 5).

For this study, we employed the Thin-ResNet (TRN) architecture~\cite{2019arXiv190210107X}, a shallow residual neural network (ResNet)~\cite{2015arXiv151203385H} with 33 layers. To accommodate the specific image input shape of the LST, the initial layer of the original TRN design was omitted. The model uses residual connections adding the input to the output at each stage, which enables deeper architectures by mitigating vanishing gradient issues. Each residual block incorporates a dual squeeze-and-excitation attention mechanism~\cite{2017arXiv170901507H}, allowing the network to emphasize the most informative features. Input images were pre-processed onto a Cartesian grid using bilinear interpolation from the \texttt{ImageMapper} module of the \texttt{DL1-Data-Handler}\footnote{\href{https://github.com/cta-observatory/dl1-data-handler}{https://github.com/cta-observatory/dl1-data-handler}}~\cite{tjark_miener_2025_15422957}, a package tailored for managing IACT data in DL workflows, enabling the use of standard convolutional layers~\cite{Nieto:2019uj}.


\section{Results}

This section presents the performance evaluation of the CNN-based analysis using the TRN, applied to simulated waveform data from the LST-1 prototype. The goal is to assess the capability of the DL model to reconstruct and classify events using calibrated waveforms. We restrict the analysis to events in which the image contains more than 50 photoelectrons (PE) after the cleaning procedure, ensuring sufficient signal quality for accurate reconstruction.

\begin{figure}[h]
    \centering
    \begin{overpic}[width=7.4cm]{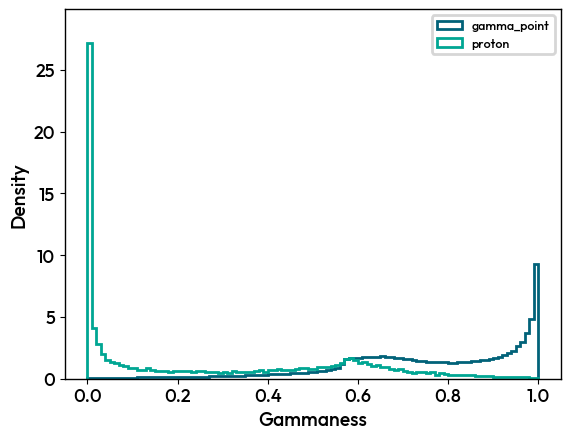} 
        \put(45,50){\textcolor{gray}{Preliminary}}
    \end{overpic}
    \begin{overpic}[width=7.4cm]{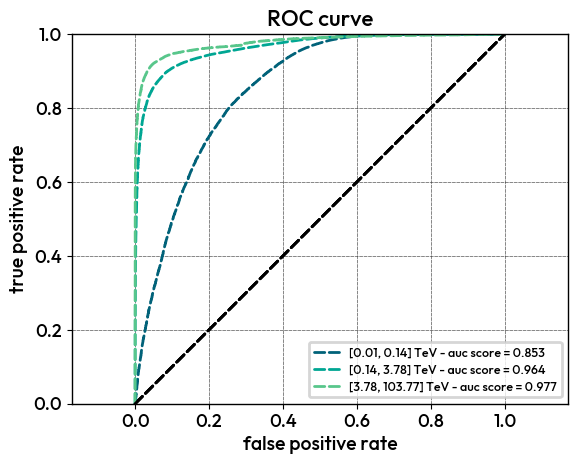} 
        \put(60,30){\textcolor{gray}{Preliminary}}
    \end{overpic}
    \caption{(\textbf{Left}): Gammaness distribution for the testing pointlike gamma sample. Only events with more than 50 PE are considered. The bump located at around 0.6 gammaness value is caused by faint events, which can be removed by increasing the threshold of PE (see also~\cite{Lacave:2025}). (\textbf{Right}): ROC curves for three different energy bins, illustrating the classifier performance across energies. The AUC score for each bin is shown in the legend, demonstrating improved separation power at higher energies.}
    \label{fig:particle_classification}
\end{figure}

\begin{figure}[h]
    \centering
    \begin{overpic}[width=4.9cm]{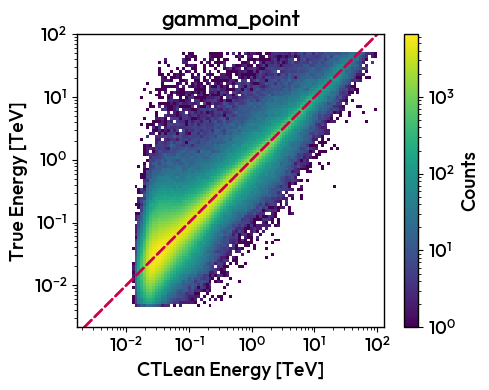} 
        \put(40,20){\textcolor{gray}{Preliminary}}
    \end{overpic}
    \begin{overpic}[width=4.9cm]{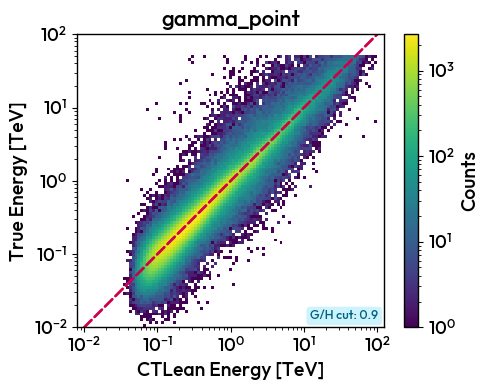} 
        \put(40,20){\textcolor{gray}{Preliminary}}
    \end{overpic}
    \begin{overpic}[width=4.9cm]{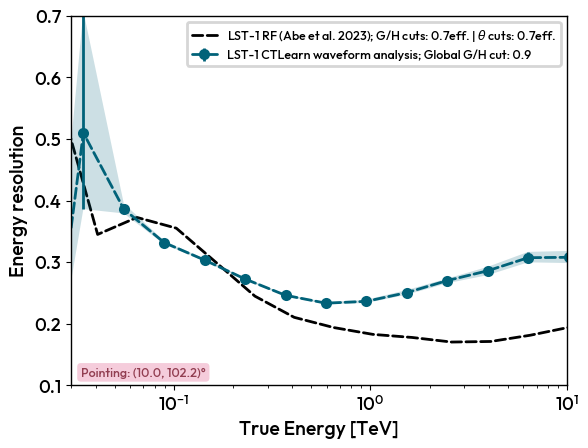} 
        \put(45,50){\textcolor{gray}{Preliminary}}
    \end{overpic}
    \caption{(\textbf{Left}): Energy migration matrix for pointlike gamma rays including all events with more than 50 PE. (\textbf{Middle}): Energy migration matrix for events passing an additional global gamma/hadron separation cut of 0.9. (\textbf{Right}): Energy resolution as a function of true energy for events passing an additional global gamma/hadron separation cut of 0.9. As a reference, the LST-1 performance~\cite{Abe_2023} obtained with the Random Forest method is indicated as the dashed black line. The reference curve is shown only for illustrative purposes and should not be directly compared to our results, as the optimization of the analysis cuts differs between methods.}
    \label{fig:energy}
\end{figure}

\begin{figure}[h]
    \centering
    \begin{overpic}[width=4.9cm]{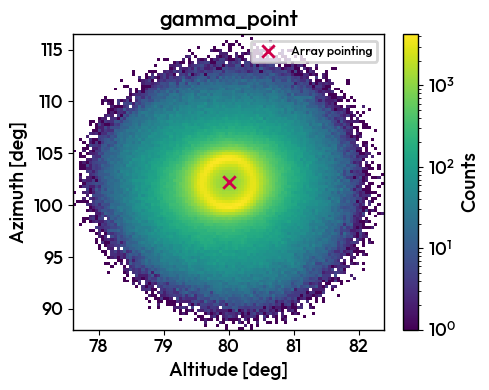} 
        \put(40,20){\textcolor{gray}{Preliminary}}
    \end{overpic}
    \begin{overpic}[width=4.9cm]{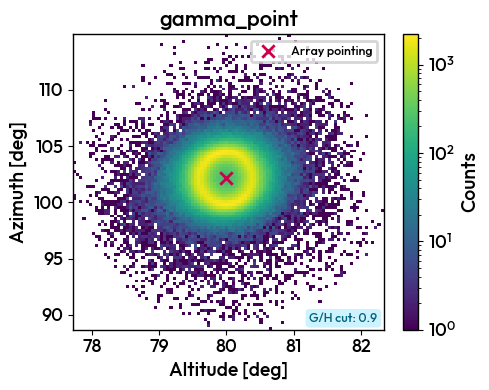} 
        \put(40,20){\textcolor{gray}{Preliminary}}
    \end{overpic}
    \begin{overpic}[width=4.9cm]{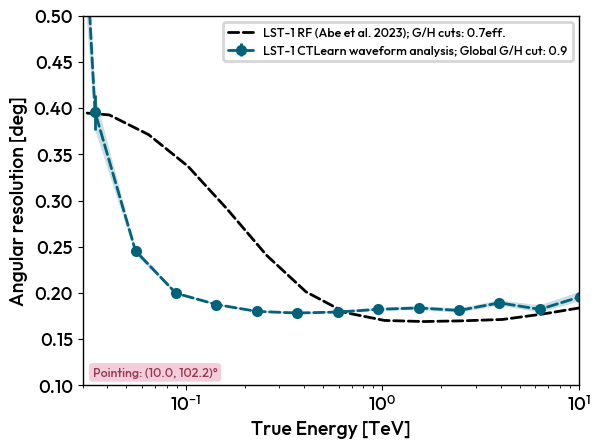} 
        \put(45,50){\textcolor{gray}{Preliminary}}
    \end{overpic}
    \caption{(\textbf{Left}): Sky map of reconstructed positions for pointlike gamma rays including all events with more than 50 PE. (\textbf{Middle}): Sky map of reconstructed positions for events passing an additional global gamma/hadron separation cut of 0.9. (\textbf{Right}): Angular resolution as a function of true energy for events passing an additional global gamma/hadron separation cut of 0.9. As a reference, the LST-1 performance~\cite{Abe_2023} obtained with the Random Forest method is indicated as the dashed black line. The reference curve is shown only for illustrative purposes and should not be directly compared to our results, as the optimization of the analysis cuts differs between methods.}
    \label{fig:angular}
\end{figure}

The classification model can be tested using the gammaness distribution for the different particle types (see left panel of Fig.~\ref{fig:particle_classification}) and the Receiver Operating Characteristic (ROC) curves together with its Area Under the Curve (AUC) score for different energy bins (see right panel of Fig.~\ref{fig:particle_classification}). The separation between gamma-ray and hadronic events is functioning as intended. Direction reconstruction accuracy is evaluated by comparing predicted arrival directions to the true ringwobble source position, which is set to an uniformly offset of the telescope pointing of $ 0.4^{\circ} $ (see left and middle panels of Fig.~\ref{fig:angular}). Energy reconstruction is assessed through energy migration matrices, showing good agreement between true and reconstructed energies for pointlike gamma rays (see left and middle panels of Fig.~\ref{fig:energy}). We then evaluate the angular and energy resolution as a function of the true energy using a global gammaness cut of 0.9 (see right panels of Fig.~\ref{fig:energy} \&~\ref{fig:angular}).

The angular resolution is defined as the angular distance within which 68\% of the reconstructed gamma-ray events fall, relative to the direction of the simulated point source of gamma rays. It is computed in logarithmic true energy bins. Similarly, the energy resolution in each true energy bin is defined as the 68\% containment of the distribution of $ (E_{\mathrm{reco}} - E_{\mathrm{true}})/E_{\mathrm{true}} $. These results serve as a first step toward validating the applicability of CNN-based models for LST observations using waveform data as input.

\section{Discussions and conclusions}

This work presents a proof-of-concept demonstrating for the first time that full event reconstruction (particle classification, energy estimation, and directional regression) can be achieved using IACT waveform data as input to CNNs. By leveraging the temporal structure contained in calibrated waveforms, our approach bypasses the need for traditional image parameterization, marking a significant step forward in the application of DL to low-level data products in the analysis chain of IACTs. However, due to the computational complexity and the time-intensive training process, spanning several weeks on high-performance hardware, we do not propose this waveform-based method as a replacement for the standard data processing pipeline used in routine IACT observations. Instead, we see its value in targeted analyses of scientifically significant or rare astronomical events, where a potential performance gain can justify the higher computational cost.

\begin{figure}
    \centering
    \includegraphics[width=13cm]{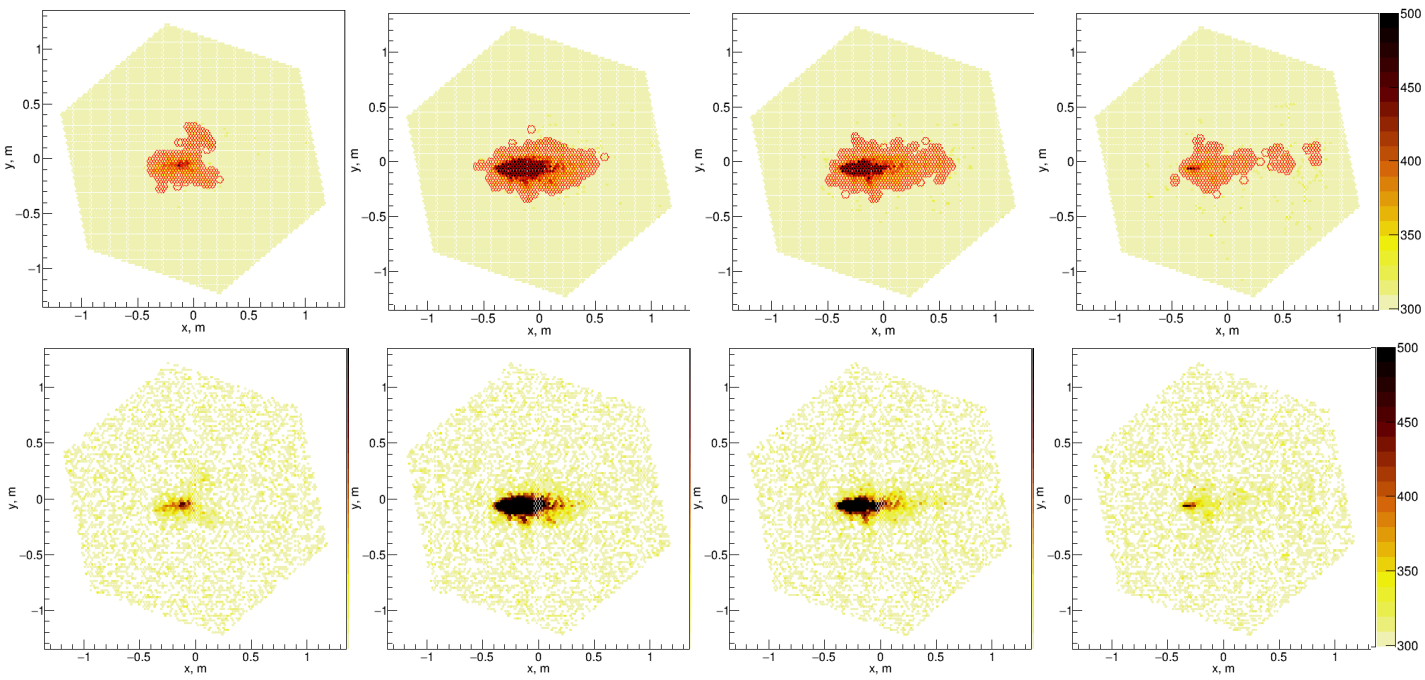}
    \caption{Four raw waveform samples with 3 nanosecond intervals for a bright $ 1.4~\text{TeV} $ gamma-ray event. The bottom panels show waveforms including night sky background noise, while the top panels display the pure Cherenkov signal. The red overlaid pixels indicate the mask detected by the DBSCAN-based cleaning algorithm at each waveform sample. The colorbar represents signal amplitude in ADC (Analog-to-Digital Converter) units.}
    \label{fig:brightgamma}
\end{figure}

\begin{figure}
    \centering
    \includegraphics[width=13cm]{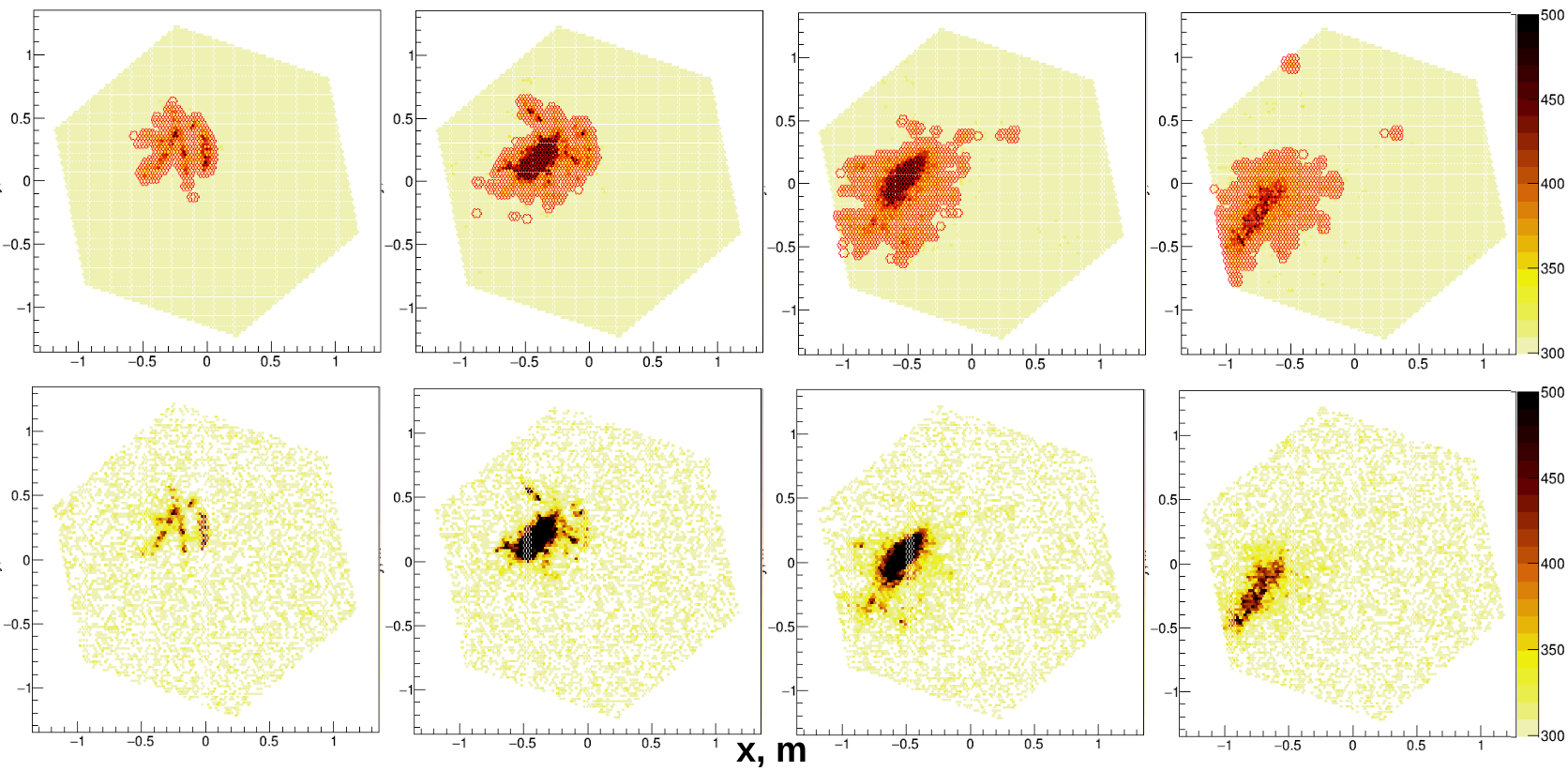}
    \caption{Same as Fig.~\ref{fig:brightgamma}, but for a bright $ 1.9~\text{TeV} $ proton event. Unlike the narrow, compact development of gamma-ray showers, the proton shower produces a much wider and more diffuse signal, reflecting the more complex hadronic interactions in the atmosphere.}
    \label{fig:brightproton}
\end{figure}

Despite the absence of a fair, one-to-one comparison, primarily due to differing optimization procedures, our proof-of-concept demonstrates that the performance achieved lies within the same bulk region as the conventional Random Forest-based analysis~\cite{Abe_2023}. This indicates that our method holds strong potential and warrants further investigation, particularly with harmonized optimization strategies that would allow a more rigorous and direct comparison.

\begin{figure}
    \centering
    \includegraphics[width=13cm]{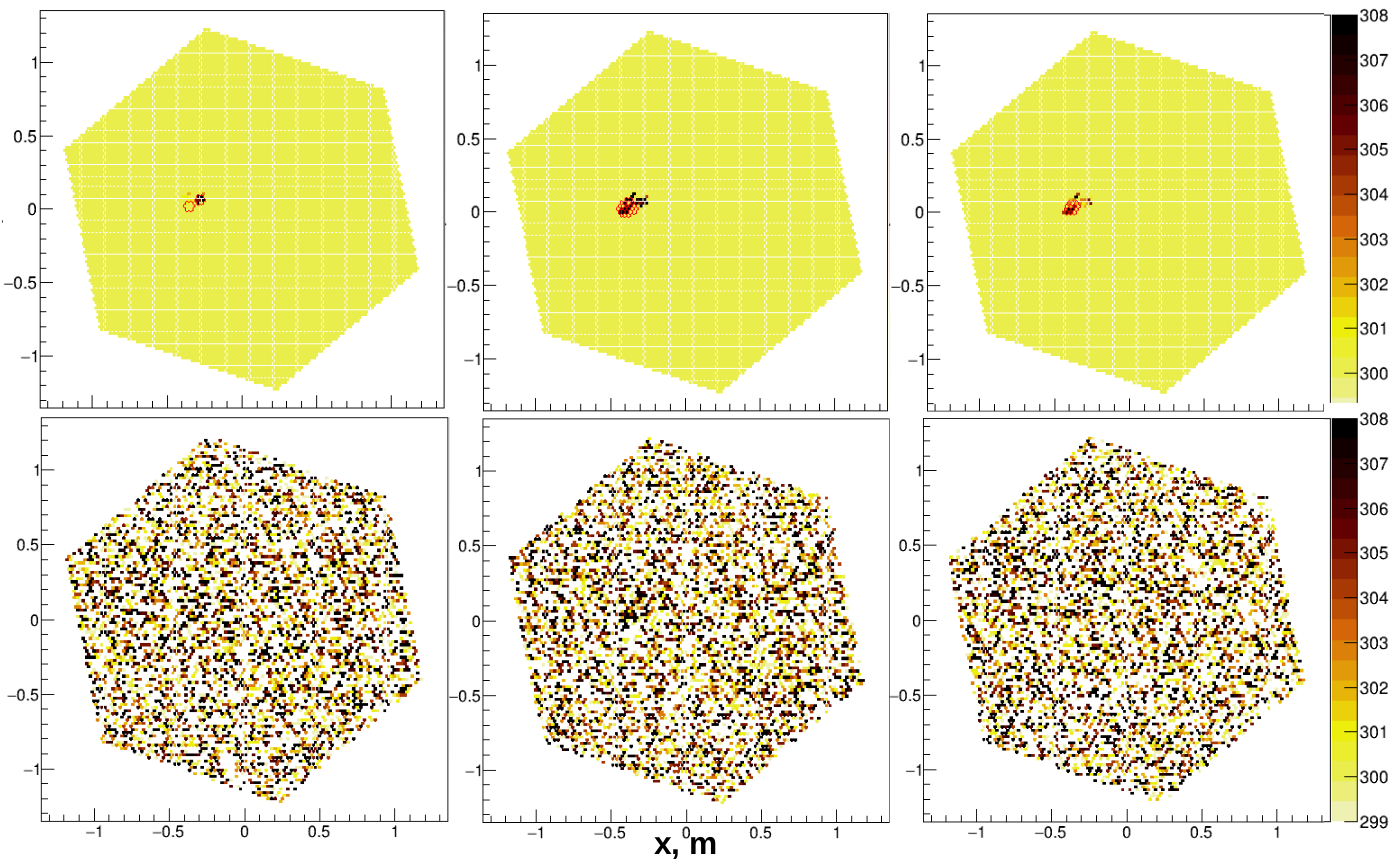}
    \caption{Three raw waveform samples with 2 nanosecond intervals for a faint 10 GeV gamma-ray event, where the signal is at the noise level but still captured thanks to the exploitation of temporal and spatial correlations.}
    \label{fig:faintgamma}
\end{figure}

To further advance this framework, we are developing a waveform-based DBSCAN cleaning algorithm to replace the default cleaning stage of the standard IACT analysis. This unsupervised ML technique for waveform cleaning is being designed for the next-generation IACT cameras (AdvCam for LSTs~\cite{Heller:2025} but is also applicable to current-generation IACT data. Figures~\ref{fig:brightgamma} through~\ref{fig:faintgamma} showcase preliminary results: Fig.~\ref{fig:brightgamma} highlights a bright gamma-ray event, Fig.~\ref{fig:brightproton} a bright proton, and Fig.~\ref{fig:faintgamma} a faint gamma-ray event of $ 10~\text{GeV} $. These examples demonstrate the algorithm’s ability to exploit temporal-spatial correlations in the waveform data, enabling the detection of faint events at noise levels that would likely be suppressed by conventional cleaning methods.

This waveform-based reconstruction method also opens the door for implementing advanced AI-driven trigger systems, as proposed to be integrated for the next-generation AdvCam cameras~\cite{Burmistrov:2025}. By directly analyzing the raw, time-resolved signals with DL models, it becomes feasible to identify and classify events in real time with high accuracy, potentially improving trigger efficiency and reducing background rates. Such AI-trigger systems could significantly enhance the sensitivity and responsiveness of future IACT arrays, making this approach a promising direction for the evolution of gamma-ray astronomy instrumentation.

\newpage

\tiny
\bibliographystyle{unsrturl}
\bibliography{biblio.bib}

\textbf{Full Author List: CTAO-LST Project}

\tiny{\noindent
K.~Abe$^{1}$,
S.~Abe$^{2}$,
A.~Abhishek$^{3}$,
F.~Acero$^{4,5}$,
A.~Aguasca-Cabot$^{6}$,
I.~Agudo$^{7}$,
C.~Alispach$^{8}$,
D.~Ambrosino$^{9}$,
F.~Ambrosino$^{10}$,
L.~A.~Antonelli$^{10}$
% ...
The CTAO-LST Project
}\\

\tiny{\noindent$^{1}${Department of Physics, Tokai University, 4-1-1, Kita-Kaname, Hiratsuka, Kanagawa 259-1292, Japan}.
$^{2}${Institute for Cosmic Ray Research, University of Tokyo, 5-1-5, Kashiwa-no-ha, Kashiwa, Chiba 277-8582, Japan}.
$^{3}${INFN and Università degli Studi di Siena, Dipartimento di Scienze Fisiche, della Terra e dell'Ambiente (DSFTA), Sezione di Fisica, Via Roma 56, 53100 Siena, Italy}.
$^{4}${Université Paris-Saclay, Université Paris Cité, CEA, CNRS, AIM, F-91191 Gif-sur-Yvette Cedex, France}.
$^{5}${FSLAC IRL 2009, CNRS/IAC, La Laguna, Tenerife, Spain}.
$^{6}${Departament de Física Quàntica i Astrofísica, Institut de Ciències del Cosmos, Universitat de Barcelona, IEEC-UB, Martí i Franquès, 1, 08028, Barcelona, Spain}.
$^{7}${Instituto de Astrofísica de Andalucía-CSIC, Glorieta de la Astronomía s/n, 18008, Granada, Spain}.
$^{8}${Department of Astronomy, University of Geneva, Chemin d'Ecogia 16, CH-1290 Versoix, Switzerland}.
$^{9}${INFN Sezione di Napoli, Via Cintia, ed. G, 80126 Napoli, Italy}.
$^{10}${INAF - Osservatorio Astronomico di Roma, Via di Frascati 33, 00040, Monteporzio Catone, Italy}.
% ...
}

\acknowledgments 
\tiny{
We gratefully acknowledge financial support from the following agencies and organisations:
Conselho Nacional de Desenvolvimento Cient\'{\i}fico e Tecnol\'{o}gico (CNPq), Funda\c{c}\~{a}o de Amparo \`{a} Pesquisa do Estado do Rio de Janeiro (FAPERJ), Funda\c{c}\~{a}o de Amparo \`{a} Pesquisa do Estado de S\~{a}o Paulo (FAPESP), Funda\c{c}\~{a}o de Apoio \`{a} Ci\^encia, Tecnologia e Inova\c{c}\~{a}o do Paran\'a - Funda\c{c}\~{a}o Arauc\'aria, Ministry of Science, Technology, Innovations and Communications (MCTIC), Brasil;
Ministry of Education and Science, National RI Roadmap Project DO1-153/28.08.2018, Bulgaria;
Croatian Science Foundation (HrZZ) Project IP-2022-10-4595, Rudjer Boskovic Institute, University of Osijek, University of Rijeka, University of Split, Faculty of Electrical Engineering, Mechanical Engineering and Naval Architecture, University of Zagreb, Faculty of Electrical Engineering and Computing, Croatia;
Ministry of Education, Youth and Sports, MEYS  LM2023047, EU/MEYS CZ.02.1.01/0.0/0.0/16\_013/0001403, CZ.02.1.01/0.0/0.0/18\_046/0016007, CZ.02.1.01/0.0/0.0/16\_019/0000754, CZ.02.01.01/00/22\_008/0004632 and CZ.02.01.01/00/23\_015/0008197 Czech Republic;
CNRS-IN2P3, the French Programme d’investissements d’avenir and the Enigmass Labex, 
This work has been done thanks to the facilities offered by the Univ. Savoie Mont Blanc - CNRS/IN2P3 MUST computing center, France;
Max Planck Society, German Bundesministerium f{\"u}r Bildung und Forschung (Verbundforschung / ErUM), Deutsche Forschungsgemeinschaft (SFBs 876 and 1491), Germany;
Istituto Nazionale di Astrofisica (INAF), Istituto Nazionale di Fisica Nucleare (INFN), Italian Ministry for University and Research (MUR), and the financial support from the European Union -- Next Generation EU under the project IR0000012 - CTA+ (CUP C53C22000430006), announcement N.3264 on 28/12/2021: ``Rafforzamento e creazione di IR nell’ambito del Piano Nazionale di Ripresa e Resilienza (PNRR)'';
ICRR, University of Tokyo, JSPS, MEXT, Japan;
JST SPRING - JPMJSP2108;
Narodowe Centrum Nauki, grant number 2023/50/A/ST9/00254, Poland;
The Spanish groups acknowledge the Spanish Ministry of Science and Innovation and the Spanish Research State Agency (AEI) through the government budget lines
PGE2022/28.06.000X.711.04,
28.06.000X.411.01 and 28.06.000X.711.04 of PGE 2023, 2024 and 2025,
and grants PID2019-104114RB-C31,  PID2019-107847RB-C44, PID2019-104114RB-C32, PID2019-105510GB-C31, PID2019-104114RB-C33, PID2019-107847RB-C43, PID2019-107847RB-C42, PID2019-107988GB-C22, PID2021-124581OB-I00, PID2021-125331NB-I00, PID2022-136828NB-C41, PID2022-137810NB-C22, PID2022-138172NB-C41, PID2022-138172NB-C42, PID2022-138172NB-C43, PID2022-139117NB-C41, PID2022-139117NB-C42, PID2022-139117NB-C43, PID2022-139117NB-C44, PID2022-136828NB-C42, PDC2023-145839-I00 funded by the Spanish MCIN/AEI/10.13039/501100011033 and “and by ERDF/EU and NextGenerationEU PRTR; the "Centro de Excelencia Severo Ochoa" program through grants no. CEX2019-000920-S, CEX2020-001007-S, CEX2021-001131-S; the "Unidad de Excelencia Mar\'ia de Maeztu" program through grants no. CEX2019-000918-M, CEX2020-001058-M; the "Ram\'on y Cajal" program through grants RYC2021-032991-I  funded by MICIN/AEI/10.13039/501100011033 and the European Union “NextGenerationEU”/PRTR and RYC2020-028639-I; the "Juan de la Cierva-Incorporaci\'on" program through grant no. IJC2019-040315-I and "Juan de la Cierva-formaci\'on"' through grant JDC2022-049705-I. They also acknowledge the "Atracci\'on de Talento" program of Comunidad de Madrid through grant no. 2019-T2/TIC-12900; the project "Tecnolog\'ias avanzadas para la exploraci\'on del universo y sus componentes" (PR47/21 TAU), funded by Comunidad de Madrid, by the Recovery, Transformation and Resilience Plan from the Spanish State, and by NextGenerationEU from the European Union through the Recovery and Resilience Facility; “MAD4SPACE: Desarrollo de tecnolog\'ias habilitadoras para estudios del espacio en la Comunidad de Madrid" (TEC-2024/TEC-182) project funded by Comunidad de Madrid; the La Caixa Banking Foundation, grant no. LCF/BQ/PI21/11830030; Junta de Andaluc\'ia under Plan Complementario de I+D+I (Ref. AST22\_0001) and Plan Andaluz de Investigaci\'on, Desarrollo e Innovaci\'on as research group FQM-322; Project ref. AST22\_00001\_9 with funding from NextGenerationEU funds; the “Ministerio de Ciencia, Innovaci\'on y Universidades”  and its “Plan de Recuperaci\'on, Transformaci\'on y Resiliencia”; “Consejer\'ia de Universidad, Investigaci\'on e Innovaci\'on” of the regional government of Andaluc\'ia and “Consejo Superior de Investigaciones Cient\'ificas”, Grant CNS2023-144504 funded by MICIU/AEI/10.13039/501100011033 and by the European Union NextGenerationEU/PRTR,  the European Union's Recovery and Resilience Facility-Next Generation, in the framework of the General Invitation of the Spanish Government’s public business entity Red.es to participate in talent attraction and retention programmes within Investment 4 of Component 19 of the Recovery, Transformation and Resilience Plan; Junta de Andaluc\'{\i}a under Plan Complementario de I+D+I (Ref. AST22\_00001), Plan Andaluz de Investigaci\'on, Desarrollo e Innovación (Ref. FQM-322). ``Programa Operativo de Crecimiento Inteligente" FEDER 2014-2020 (Ref.~ESFRI-2017-IAC-12), Ministerio de Ciencia e Innovaci\'on, 15\% co-financed by Consejer\'ia de Econom\'ia, Industria, Comercio y Conocimiento del Gobierno de Canarias; the "CERCA" program and the grants 2021SGR00426 and 2021SGR00679, all funded by the Generalitat de Catalunya; and the European Union's NextGenerationEU (PRTR-C17.I1). This research used the computing and storage resources provided by the Port d’Informaci\'o Cient\'ifica (PIC) data center.
State Secretariat for Education, Research and Innovation (SERI) and Swiss National Science Foundation (SNSF), Switzerland;
The research leading to these results has received funding from the European Union's Seventh Framework Programme (FP7/2007-2013) under grant agreements No~262053 and No~317446;
This project is receiving funding from the European Union's Horizon 2020 research and innovation programs under agreement No~676134;
ESCAPE - The European Science Cluster of Astronomy \& Particle Physics ESFRI Research Infrastructures has received funding from the European Union’s Horizon 2020 research and innovation programme under Grant Agreement no. 824064.}

\end{document}